\documentclass[a4,paper,11pt]{article}
\usepackage[left]{lineno}
\headheight\baselineskip \headsep2\baselineskip \textwidth210mm
\advance\textwidth -2in \textheight287mm \advance\textheight-2in
\usepackage{epsfig}
\usepackage{graphicx}
\usepackage[colorlinks]{hyperref}
\usepackage{epstopdf}
\label{sec.sub.gdn}

\newtheorem{Theorem}{Theorem}

\usepackage{amsfonts}
\usepackage{amsmath}

\begin{document}
\title{Sampling of multiple variables based on partial order set theory}
\author {\textbf{Bardia Panahbehagh}\\ Department of Mathematics and computer science, Kharazmi University,\\ Tehran, Iran, panahbehagh@khu.ac.ir; bardia.panah@gmail.com\\\\   \textbf{Rainer Bruggemann}\\ Leibniz-Institute of Freshwater Ecology and Inland Fisheries,\\ Berlin, Germany, brg\_home@web.de\\\\
\textbf{Mohammad M. Salehi}\\ Department of Mathematics,
Statistics and Physics, Qatar University,\\ P.O. Box 2713, Doha,
Qatar, salehi@qu.edu.qa }

\maketitle

\begin{abstract}
This paper is going to introduce a new method for ranked set
sampling with multiple criteria. The method is based on a version of ranked set sampling, introduced by Panahbehagh et al. (2017), which relaxes the restriction of selecting just one individual variable from each ranked set. Under the
new method for ranking, elements are ranked in sets based on
linear extensions in partial order sets theory, where based on all the
variables simultaneously. Results will be evaluated by some simulations and two real case study on economical, medicinal use of flowers and the pollution of herb-layer by Lead,
 Cadmium, Zinc and Sulfur in regions in the southwest of Germany.
\end{abstract}

\section{Introduction}
Ranked set sampling (RSS) was first introduced by McIntyre
(1952) and has been widely used as a design in many applications.
The idea behind RSS is appealing particularly to agricultural
and environmental scientists where identifying sampling units in
the field is straightforward but the exact exploration measurement of the units by measurements is
time consuming. Many sampling units can be identified and within
them  subsets are actually measured. In RSS the identification
of these subsets is based on ranking the units and a selection
according to their relative ranks.

The RSS technique briefly involves taking random samples of size
$m$ from the population. The sample units are ranked by some quick
and easy measure. Then, one unit from each sample is chosen and
precisely measured  for the character of interest. To take a
sample of size $m$, the unit that has the lowest rank in the first
sample (with size $m$) is chosen, the unit with the second lowest
rank is chosen from the second sample, and so on. This process is
repeated $n$ times, giving a final sample size, $n_.=nm$. Sampling can
be balanced or unbalanced where the number of sample units
selected in the ranks are not constant. With highly skewed
population distributions more units from low (or high) ranks can
be selected. Unbalanced designs are similar in concept to the
optimal allocation in stratified sampling where strata with bigger
variances, take bigger sample fractions. RSS is reported as being
more efficient than simple random sampling (Ridout, 2003;
Samawi, 1996). See full reviews of RSS by
 Patil et al. (1999) and the related book of Chen et al. (2004).\\\\
In this paper, based on method of Panahbehagh et al. (2017) multivariate RSS based on partial order sets will be introduced.
In some populations there are more than one character of interest,
%for example, multiple species in a survey of forest bio-diversity.
%Various approaches of RSS with multiple variables have been
%suggested.
 Patil et al. (1994) have discussed RSS for multiple
variables when one of the variable can be defined as a primary
variable. Ranking is based on this main primary variable only, and
if the other variables are correlated with the main one, the
method will perform reasonably well. Norris et al. (1995) have
developed two approaches, one using an unbalanced allocation
process based on the Neyman allocation for the variable of
primary interest, treating this as a concomitant for the other
variables of interest and the other using a design based on
randomly choosing sample units from the rank list derived from an individual variable. Al-Saleh and Zheng (2002) as well as Chen and
Shen (2003) have proposed a two-layer ranked set sampling for the
situation in which we have two main variables or two concomitant
variables to rank the data.
% Their methods have an advantage that
%consider the two variables, each in a layer, and try to present
%the data with all required ranks.
In their methods at the first layer, the data is ranked based on the first variable and a RSS sample is selected. At the second round, the first layer RSS data will be ranked based on the second variable and the RSS data in the second layer will be present as the final sample. One disadvantage of their methods is that they
consider the two variables separately, and not simultaneously.
Another disadvantage is that they are requiring many
initial samples to achieve the needed sample size and also with
increasing dimension of the space of variables,
the size of the needed sample will increase severely.\\\\
In this paper, applying the framework developed by Panahbehagh et al. (2017) multivariate RSS based on partial order sets will be introduced.\\
We demonstrate our suggested sampling technique with two environmental examples:\\

\begin{itemize}

\item  The first example deals with the estimation of mean values of “flower dry weight” and “essence” of Matricaria chamonilla, which is considered as a very important commercial and medicinal plant in Iran and many other countries. The main part of chamomile for medicinal purposes is the flower essence and it is economically important to maximize the oil yield. It is hardly possible to measure the efficiency of oil yield under all scenarios and all suitable geographical units within Iran. Therefore sampling technique is necessary and is performed.\\\\

\item Chemical pollution in the environment is a problem which came into the focus of administration since the early eighties. Chemicals pose a hazard to humans, animals, plants, etc. due to their toxicity. The quantification of the hazard is however extremely difficult as uptake mechanisms, mode of toxic action, the role of chemical speciation and the state of the environmental geographical unit are important. Therefore in almost all nations monitoring programs were installed to observe the chemical pollution spatially and temporarily. The data have mostly the unit mass of chemicals (as total concentration) mass of the target, for example soil.\\
     These data are thought of as surrogates expressing the hazard potential due to the considered chemicals. It is difficult to obtain for example mean values of concentrations taking into account all geographical units, especially when a temporal trend is to be monitored. Here 59 geographical units are selected by the environmental protection agency taking care for defining the regions as homogeneous as possible with respect to the chemical pollution processes. The sampling technique can be validated, because in that specific case the mean values can also directly obtained from all 59 units for a specified year of observation. When the proposed method is successful then the monitoring process can be simplified, namely to relax the precondition of almost homogeneous geographical units and a more elaborated locally specific monitoring can be applied.
\end{itemize}

To develop the new method, in section 2 we extend the method of Panahbehagh et al. (2017) for multiple variables. In section 3 we
introduce stratified sampling using RSS derived from linear extensions (LE) in partial order sets (Posets). Section 4 contains examples,
simulations and two real case study to compare the methods and
evaluate the results and the paper will be finished in section 5
with a conclusion.

\section{\protect\bigskip Multivariate Virtual Stratified Ranked Set Sampling (MVSR)}
 In multivariate RSS, we have an $R$ dimensional random
variable. We start with the basic idea of multivariate RSS (Patil et al., 1994), ranking according to just one of the variables.
Then we adapt the design with the design of Panahbehagh et al. (2017).\\
Suppose that $\mathbf{X}\sim f_{\mathbf{\mu}}$ with $E(\mathbf{X}%
)=\mathbf{\mu},$ where $\mathbf{X}=(X^{1},X^{2},...,X^{R})$
and $\mathbf{\mu}=(\mu ^{1},\mu ^{2},...,\mu ^{R})$ also $%
Var(X^{j})=\sigma _{j}^{2} $, $Cov(X^{j},X^{j^{\prime }})=\rho
_{jj^{\prime }}\sigma _{j}\sigma _{j^{\prime }}$ and
$E(|X^j|^2)<\infty$ for all $j$. Main aim is to estimate
$\mathbf{\mu}$. Our strategy to get a sample of size $n_.=nm$ from
the population is to generate an $iid$ sample of $\mathbf{X}_{i}$s of size $m$ from $f$ and sort
them according to $X^{1}$ (using itself or based on an auxiliary variable) in $m$ columns
and repeat this method $K$ times.
Then we will have a stratified population, formed in $m$ strata,
each of size $K$ (see table \ref{tab2}), just assume we have a vector of $\mathbf{X}_{(h)i}$ instead of a $X_{(h)i}$, where $\mathbf{X}_{(h)i}=(X_{(h)i}^{1},X_{[h]i}^{2},...,X_{[h]i}^{R})$, and
$X_{(h)i}^{1}$ is the $ h^{th}$ order statistics in the $i^{th}$
set with $\mu _{(h)}^{1}$and $\sigma
_{(h)1}^{2}$ as the mean and variance respectively, and $X_{[h]i}^{j}$ for $%
j=2,3,..,R$ are concomitant variables with respect to $X_{(h)i}^{1}$ in $i^{th}$ set with $\mu _{[h]}^{j}$and $\sigma _{[h]j}^{2}$.
%Then for $_{(.)}$ ranking is assumed without error as based on the ranking of a leading variable, whereas in $_{[.]}$ there are errors in the ranking.
Now we get a Simple Random Sampling Without Replacement (SRSWOR) from the
$h^{th}$ stratum of size $n$ (an integer smaller than $K$), say $s_{h},$ and we
can estimate $\mu ^{j}$ by%
\begin{eqnarray*}
\widehat{\mu }_{V}^{1} &=&\frac{1}{m}\sum\limits_{h=1}^{m}\bar{X}%
_{(h)}^{1}, \\
\widehat{\mu }_{V}^{j} &=&\frac{1}{m}\sum\limits_{h=1}^{m}\bar{X}%
_{[h]}^{j}
\end{eqnarray*}

where%
\begin{eqnarray*}
\bar{X}_{(h)}^{1} &=&\frac{1}{n}\sum_{i\epsilon s_{h}}X_{(h)i}^{1}, \\
\bar{X}_{[h]}^{j} &=&\frac{1}{n}\sum_{i\epsilon
s_{h}}X_{[h]i}^{j}
\end{eqnarray*}

\begin{Theorem}\label{theo1}
In MVSR, $\widehat{\mathbf{\mu}}^j_{V}$ is an unbiased
estimator for
$\mathbf{\mu}^j$ and%
\begin{eqnarray*}
V(\widehat{\mu }_{V}^{1}) &=&\frac{1}{nm}(\sigma _{1}^{2}-%
\frac{(1-\frac{n}{K})}{m}\sum\limits_{h=1}^{m}(\mu _{(h)}^{1}-\mu
^{1})^{2}), \\
V(\widehat{\mu }_{V}^{j}) &=&\frac{1}{nm}(\sigma _{j}^{2}-%
\frac{(1-\frac{n}{K})}{m}\sum\limits_{h=1}^{m}(\mu _{\lbrack h]}^{j}-\mu
^{j})^{2})
\end{eqnarray*}

and if we assume that $X^{1}$ and $X^{j}$ are linked with below
linear regression model

\begin{equation}\label{cond}
X_{i}^{j}=\mu ^{j}+\rho _{1j}\frac{\sigma _{j}}{\sigma
_{1}}(X_{i}^{1}-\mu ^{1})+\varepsilon _{i}
\end{equation}

where $\varepsilon $ is a random variable independent  from $X^{1}$, then%
\begin{equation*}
V(\widehat{\mu }_{V}^{j})=\frac{1}{nm}(\sigma _{j}^{2}-%
\frac{(1-\frac{n}{K})}{m}\rho _{1j}^{2}\sum\limits_{h=1}^{m}(\mu
_{(h)}^{j}-\mu ^{j})^{2})
\end{equation*}
and
\begin{eqnarray*}
\widehat{V}(\widehat{\mu }^1_{\mbox{\tiny{V}}})=\frac{K-1}{m(mK-1)}\sum\limits_{h=1}^{m}\frac{1}{n(n-1)}\sum_{i\epsilon
s_{h}}(X^1_{(h)i}-\bar{X}^1_{(h)})^2+\frac{1}{m(mK-1)}\sum\limits_{h=1}^{m}(\bar{X}^1_{(h)}-\widehat{\mu}^1_{\mbox{\tiny{V}}})^2
\\
\widehat{V}(\widehat{\mu }^j_{\mbox{\tiny{V}}})=\frac{K-1}{m(mK-1)}\sum\limits_{h=1}^{m}\frac{1}{n(n-1)}\sum_{i\epsilon
s_{h}}(X^j_{[h]i}-\bar{X}^j_{[h]})^2+\frac{1}{m(mK-1)}\sum\limits_{h=1}^{m}(\bar{X}^j_{[h]}-\widehat{\mu}^j_{\mbox{\tiny{V}}})^2
\end{eqnarray*}

are unbiased estimators for the variance of variables.
\end{Theorem}
For the proof of Theorem \ref{theo1} see Appendix A.\\
As we saw, in MVSR one is selected as a leading one to perform a ranking, the others are just adjusted which implies some errors. Therefore we introduce a method of ranking that ranks all variables
simultaneously.

\section{Ranking based on Posets}
 In this section, we first describe Posets theory and then introduce two new versions of multivariate RSS, based on them.

\subsection{Posets and Linear Extensions}
The application of theory of partial orders for ranking has been
described by Bruggemann and Patil (2011). In this theory, we have a set containing $m$ elements each of them with $R$ variables,
 with a binary relation between the elements. To compare two
elements of the set, if all variables of the first element are equal or bigger (smaller) than the second one, then the first element is better ($\geq $) (worse ($<$)) than second one, otherwise the two elements
are not comparable.
Linear extensions (LEs) are different projections of the partial order into a complete
order that respect all the relations in the
partial order set. I.e. Linear extensions are the result of order preserving mappings. Therefore a relation $x<y$ in a poset is preserved in all linear extensions.\\\\
%Note, an object set $O$ equipped with a complete order means, that all pairs $(x,y)\in O\times O$ are mutually comparable. In that case a ranking can be performed. In general, with $R$ variables no complete order is obtained. Then a set of linear extensions instead of one ranking is the outcome.  \\
We use this theory to introduce two designs; Ranking based on Posets using complete form (or at least a random sample) of LEs (CPOR) and Ranking based on Posets using just one random selection of LEs (RPOR):
\begin{itemize}
   \item [CPOR:] First rank the elements according to the mean height of the elements due to all the possible LEs where height is defined as the rank of the element in the respective LE and then construct an unequal size population using these mean heights based on complete LEs.
  \item [RPOR:] Select one of the LEs to construct an equal size population.

\end{itemize}
We illustrate the topic
with an example where we assume a set with $m=5$ and $R=2$ (see
table \ref{tab4}). The set of all LEs obtained from the data in table \ref{tab4} is shown in table \ref{tab5}. Here, due to the low number of linear extensions, the average height of each element can be easily directly determined from table \ref{tab5}.\\
Generally, the  determination of all linear extensions is computationally a hard problem.
Therefore the determination of average heights needs themselves sampling techniques as shown by
Bubley and Dyer (1999). However, it is not necessary to determine the set of LEs explicitly, because only the average height is of interest. In this case, there are also pretty good approximations available, see for instance Bruggemann et al. (2004), (2013) or De Loof et al. (2013). According to the heights of each element in LEs form, we have table \ref{tab6}.\\\\
We will use above theory to stratify each set in the next
subsection.

\subsection{CPOR}
 We are going to put each element of a
set into a stratum equal to the nearest integer of the mean of its
height (MH).
Following the previous example according to table \ref{tab5}, we will put
the elements of the set into 5 virtual strata (see table \ref{tab7}).\\
Then, the design proceeds as follow: an $iid$ sample of size $m$
(a set) from $f$ will be generated, and according to their
variables ($X^{j}s$) all possible linear extensions will be
constructed. We then calculate the mean height (either explicitly by determination of the set of all LEs or directly by applying approximations). Finally, using these heights, put the elements of the set into the strata and
repeat this approach $K$ times. It is obvious that this method
leads to an unequal size stratified population.\\
Then instead of a R dimensional variable $\textbf{X}_{\{h\}i}=(X_{\{h\}i}^{1},X_{\{h\}i}^{2},...,X_{\{h\}i}^{R})$ we have a R+1 dimensional variable $\textbf{X}_{\{h\}i}=(X_{\{h\}i}^{1},X_{\{h\}i}^{2},...,X_{\{h\}i}^{R},MH_{\{h\}i})$ where $MH$ stands for the mean heights of the objects.\\
We now have a stratified population with unequal size. For the
$h^{th}$ stratum we will take a SRSWOR,  $s_{h},$ with size
$n_{h}$, proportional to the stratum size,  $K_{h}$, where
$\sum\limits_{h=1}^{m}K_{h}=Km$ such that $\sum\limits_{h=1}^{m}n_{h}=n_.=nm$. The
stratified population is presented in table \ref{tab8}.\\
In table \ref{tab8}, $\textbf{X}_{\{h\}i}=(X_{\{h\}i}^{1},X_{\{h\}i}^{2},...,X_{\{h\}i}^{R})$\, where $X_{\{h\}i}^{j}$ is the $j^{th}$ character of an element that has been fallen into the $h^{th}$ stratum after $i-1$ elements,
according to its mean height MH in respective LEs. Now we propose an
estimator for $\mu^{j}$ (the expectation of the $j^{th}$ character in $f$) as
\begin{equation*}
\widehat{\mu}_{P}^{j}=\sum\limits_{h=1}^{m}W_{h}\bar{X}_{\{h\}}^{j}
\end{equation*}

where
\begin{equation}\label{weight}
W_{h}=\frac{K_{h}}{Km}
\end{equation}
and
\begin{equation*}
\bar{X}_{\{h\}}^{j}=\frac{1}{n_{h}}%
\sum_{i\epsilon s_{h}}X_{\{h\}i}^{j}
\end{equation*}

\begin{Theorem}\label{theo2}
In CPOR, $\widehat{\mu}_{P}^{j}$ is an unbiased estimator
for $\mu^{j}$.
\end{Theorem}
For the proof of Theorem \ref{theo2} see Appendix B.\\
%\begin{Proof}
%  With

%\begin{equation*}
%\widehat{\mu}_{P}^{j}=\sum\limits_{h=1}^{m}W_{h}\bar{X}_{\{h\}}^{j}
%\end{equation*}
%where%
%\begin{eqnarray*}
%I_{\{h\}i}=\left\{\begin{matrix}
%1
% & \;\;\;\;  \text{if $X^{j}_{\{h\}i}$ is in the $s_{\{h\}}$}
% &\\\\
%0 & \;\;\;\;  \text{otherwise}
% &
%\end{matrix}\right.
%\end{eqnarray*}

%we have
%\begin{eqnarray*}
%E(\widehat{\mu }_{P}^{j}) &=&E_{M}[E_{D}(\widehat{\mu }_{P}^{j}|%
%\mathbf{X}_{Km})]=E_{M}[\sum\limits_{h=1}^{m}\frac{K_{h}}{Km}\frac{1}{n_{h}}%
%\sum_{i=1}^{K_{h}}X_{\{h\}i}^{j}E_{D}(I_{\{h\}i}|\mathbf{X}_{Km})] \\
%&=&E_{M}(\sum\limits_{h=1}^{m}\frac{K_{h}}{Km}\frac{1}{n_{h}}\sum_{i=1}^{K_{h}}X_{%
%\{h\}i}^{j}\frac{n_{h}}{K_{h}})=\frac{1}{Km}E_{M}(\sum\limits_{h=1}^{m}%
%\sum\limits_{i=1}^{K}X_{hi}^{j})=\mu^{j}
%\end{eqnarray*}
%\end{Proof}
%\textcolor{red}{For variance of the estimator we have?????? I am not sure, because conditional expectation of $E(X^j_h|K_h)=\sigma_h?!?!?\sigma_{K_h}$:}
%\begin{equation*}
%Var(\widehat{\mu}_{P}^{j})=\frac{\sigma^2_j}{Km}+\sum\limits_{h=1}^{m}E(\frac{K_{h}^2}{(Km)^2}\frac{1-\frac{n_h}{K_h}}{n_h}S^2_{\{h\}jK_h})
%\end{equation*}

Here instead of Neyman allocation, proportional to size is used
that is easy to implement and does not need extra information (Sarndal et al. 1992).
\subsection{RPOR}
RPOR is easier than CPOR to perform. Here it is just enough to select (or construct) on of the LEs in table \ref{tab5} randomly and put them in 5 strata and then we will have a stratified population, formed in $m$ strata,
each of size $K$ like MVSR (see table \ref{tab2}). Here we show the vector of $i^{th}$ variable in $h^{th}$ stratum with $\textbf{X}_{[h\}i}=(X_{[h\}i}^{1},X_{[h\}i}^{2},...,X_{[h\}i}^{R})$. Now we get a SRSWOR from the $h^{th}$ stratum of size $n$ (an integer smaller than $K$), say $s_h$. Now we propose an
estimator for $\mu^{j}$ as
\begin{equation*}
\widehat{\mu}_{R}^{j}=\frac{1}{m}\sum\limits_{h=1}^{m}\bar{X}_{[h\}}^{j}
\end{equation*}
where
\begin{equation*}
\bar{X}_{[h\}}^{j}=\frac{1}{n}\sum\limits_{i\in s_h}X_{[h\}i}^{j}
\end{equation*}

\begin{Theorem}\label{theo3}
In RPOR, $\widehat{\mu}_{R}^{j}$ is an unbiased estimator
for $\mu^{j}$ with variance
\begin{equation}\label{VR}
V(\widehat{\mu}^j_{R})=\frac{\sigma^2_j}{Km}+\frac{1}{m^2}\sum\limits_{h=1}^m\frac{1-\frac{n}{K}}{n}E_M(\frac{1}{Q}\sum\limits_{q=1}^QS^2_{[h\}qjK}).
\end{equation}
where $q=1,2,...,Q$ are all the possible combinations of LEs, with the below unbiased estimator of variance
\begin{eqnarray*}
\widehat{V}(\widehat{\mu}^j_{R})=\frac{1}{nm(Km-1)}[\sum\limits_{h=1}^{m}\sum_{i\epsilon s_{[h\}}}(X^j_{[h\}i}-\widehat{\mu}_{R}^{j})^2+(K-n)\sum\limits_{h=1}^{m}s^2_{[h\}j}].
\end{eqnarray*}
where $S^2_{[h\}qjK}$ and $s^2_{[h\}j}$ are variance of $h^{th}$ stratum under $q^{th}$ combination of LEs and sample variance of $h^{th}$ stratum for $j^{th}$ variable respectively.
\end{Theorem}
For the proof of Theorem \ref{theo3} see Appendix C.\\
\subsection{Negative Correlation}\label{nagative}
When correlation between variables are strongly negative,
according to Posets theory, it is probable that most of the
elements in a set are incomparable. This can make it
meaningless to stratify the sets (note that in this case most of
the elements will fall in the middle stratum).\\
An extreme case is when   the correlation between two variables is
"-1". All the generated elements will be incomparable and  in the
LEs the mean height of all of them will be the same and all will
fall in the same stratum. The weight of the stratum (equation (\ref{weight})) will be 1 (and
the other strata zero). Finally we will take a
simple random sampling without replacement of size $n_.=mn$ from the stratum and the design will essentially become simple random sampling with replacement.\\
To overcome this problem, we suggest that if the bivariate correlations
between some variables are negative, multiple a "-1" to some of
them to change the correlations to positive. But if we have more
than two variables, sometimes it would not possible to make all
the correlations positive. In such cases, it is better to select
some more important variables that we are able to make their
correlations positive. We then rank the elements using Posets
theory with this new correlations.\\
In Bruggemann and Patil (2011) a procedure is explained, how subsets of variables can be systematically found. The crucial concept is the number of incomparabilities of a poset. First a sensitivity measure for each variable is to be defined. The sensitivity measures the impact of each variable on the structure of the poset (roughly: the system of comparabilities within a poset). Secondly  the variables are ordered due to their impact on a poset. Thirdly considering first the poset, due to the most sensitive variable, then the poset, due to the first two most important variables, etc the number of incomparabilities is calculated as function of the merged variables. The resulting curve motivates to find subsets of variables, which constitute mainly the poset. The remaining variables are considered as fine tuning, and will be ignored.\\
 % To estimate the parameters, the original data should be used.

\section{Simulation Study}
To evaluate and compare the efficiency of the designs, we
calculate
\begin{equation*}
\mbox{Efficiency}(\widehat \mu_.)={\\V(\bar y)\over \mbox{MSE}(\widehat \mu_.)}
\end{equation*}
where $\bar y$ is the sample mean of a simple random sample, and $\widehat \mu_.$
stands for $\widehat \mu_{V}$ (MVSR design), $\widehat \mu_{P}$ (CPOR design) or $\widehat \mu_{R}$ (RPOR design) and MSE indicate mean square error.\\
This section contains 3 parts:
\begin{itemize}
\item Comparing CPOR and RPOR with MVSR using
some simulations \item Comparing CPOR and RPOR with MVSR using a real
case study on medical flowers
\item Comparing CPOR and RPOR with MVSR using a real
case study on environmental pollution.

\end{itemize}
Also in the simulations, no matter how small was size or variance of a particular stratum, at least one sample is dedicated to the stratum. All the simulations are done by "R 3.1.2" software. For the Monte Carlo simulation we have used 20000 iterations. Expectations, variances and MSEs of the estimators are computed using Mote Carlo method.

\subsection{Comparing CPOR and RPOR with MVSR using some simulations}
In this part we will investigate efficiency of the designs that
are introduced in section 2 and 3, using bivariate normal distribution (with solving
negative correlation problem).\\

\subsubsection{Bivariate Normal distribution with negative correlation }
Here we performed the simulation assuming normal distribution with negative
correlation, with $n=4,K=8$ and $m=3$. As we can see in table \ref{tab12},
and as we asserted in Section \ref{nagative}, when the correlation is
strongly negative, CPOR and RPOR decline to simple random sampling (efficiency$\simeq
1$). When we convert the correlation to a positive value by
changing the sign of one variable, the efficiency problem will be
solved (compare the results in the last two columns with the
results in the first two columns).\\

\subsubsection{Bivariate Normal}
More complete simulations for bivariate Normal distribution are shown in
table \ref{tab13}.
For all the cases we simulated bivariate normal with $\mu^1=0$, $\mu^2=0$, $\sigma_1=1$, $\sigma_2=1$ and $\rho=0.3, 0.5, 0.7, 0.9$.\\
 First note that changing $\rho$, does not affect the efficiency of the first variable which is confirmed by simulations with less than 0.02 error. As a general point, CPOR and RPOR designs
increase the efficiency of the estimator for both variables,
simultaneously, whereas the traditional multivariate ranked set
sampling just enhances estimation of one of the variables. As the
correlations increase, efficiency increase. Unlike MVSR, CPOR and RPOR
had good and reasonable efficiency with all the correlations. Also CPOR that uses all information of LEs was more efficient than RPOR.\\
\subsection{Comparing CPOR and RPOR with MVSR using a real
case study on medical flowers}
To evaluate the designs in this section we used a real case study data on chamomile flower (Panahbehagh et al. 2017) as an medicinal use of flowers. We consider the population mean of the "Flower dry weight" (Fdw)
and "Essence" (Esn) as the two main parameters. Because we have no
information about them before sampling, and it is expensive to
measure them, we used two auxiliary variables, easy to measure
with reasonable correlation with the two main variables. For
sorting Fdw, we used "Flower height" (Fht) with correlation
of 0.78 and for Esn we used "Number of petals" (Npt) with correlation
of 0.71. Also the correlation between Fht and Npt was 0.77. Simulation results are in table \ref{tab14}. As we can see in
table \ref{tab14}, CPOR and RPOR enhance efficiency of both of the estimators
simultaneously. The most important factor in efficiency is
the portion of $K/n$ and efficiency increased with increasing this factor. For example compare two cases: one $m=5,
K=7, n=3$ and two $m=5, K=7, n=5$,  although $n$ is larger in the
second case, because the portion of $K/n$  is larger for first
one, the efficiency of the first case is larger than the second
case. Also if the other parameters are equal, $m$
is the other important parameter that affect efficiency and efficiency increased with increasing $m$. Again CPOR was more efficient than RPOR in almost all the cases.\\
%The results show that MVSR has better performance for "Ffw" in
%all the cases because the first variable uses $\rho=1$, as MVSR
%just uses the first variable for sorting and sampling, and CPOR
%sacrifices some efficiency for "Ffw" and enhances the efficiency
%of "Esn", as we expected. The most important factor for efficiency
%of the estimators is also the proportion of "K/n". Also the results for "Esn" as the second variable, are different between left and right side of table \ref{tab14} because of different variables that are used as the first variable (Fdw and Ffw). \\

\subsection{Chemical Pollution}
The Environmental Protection Agency (EPA) of the German state Baden-Wuerttemberg performed a series of measurements in different targets, for example in the herb layer, in the epiphytic mosses of trees, in fish etc.. For this purpose the state Baden Wuerttemberg was divided in 60 more or less homogenous regions with respect to their natural environment. The regions are not selected according to administrative classification but to get regions as homogeneous as possible with respect to environmental pollution processes.\\
The task was and is, to protocol the pollution due to industry, traffic, agrarian management with respect to the total concentrations of Lead, Cadmium, Zinc and Sulfur (measured in mg/kg dry mass).\\
According to the different emission types there are different chemical species, for example SO$_2$ or solved in atmospheric droplets H$_2$SO$_3$, similarly the other metals as for example Pb, which can be bounded in organic chemicals or as oxids.\\
The different targets, selected by the EPA should help to differentiate among the different transport processes and to be able to trace back the emission source. So, the herb layer is mainly a short range transport indicator, whereas the epiphytic mosses (simply: moss layer) is considered as indicating middle range transports. The herb layer should especially indicate the loading due to the public traffic whereas the moss layer may mainly indicate industrial sources.\\
An interesting point of geochemical research is as to how far the presence of e.g. Pb implies the presence of Cadmium. A first attempt in this direction can be found in a paper by Bruggemann, Kerber, 2018 (submitted to a special issue of Comm.in Math. and in Comp. Chemistry). A classification approach concerning the pollution of Baden - Wuerttemberg was published by Bruggemann et al. (2013).

\subsubsection{ Comparing CPOR and RPOR with MVSR using a real
case study on environmental pollution }
In this study, regions in Baden-Wuerttemberg, South-West of Germany were selected and
monitored with respect to total concentrations of the chemical elements Pb, Cd, Zn and S in the
herb layer (Environmental Protection Agency Baden-Wurttemberg (Germany) 1994, Signale aus der Natur). The herb layer is one of the targets, selected by the Environmental Protection Agency of Baden-Wuerttemberg. This multi-indicator system with regions as objects and concentrations of the four
chemical elements as indicators (Bruggemann and Patil 2011) raises the questions:
\begin{itemize}
\item How can we get information about the pollution status?
\item What can be said about geochemical relations?
\end{itemize}
For example does an increase in pollution with respect to one pollutant,for example Pb, always imply the
increase of another pollutant, for instance Cd? For an answer from the point of view of applied partial order theory, see Bruggemann and Voigt (2012) (For more details see Bruggemann et al., 1996; Bruggemann et al., 1998; Bruggemann et al., 1999; Bruggemann et al., 2003 and Bruggemann et al., 2013). \\\\
Here to give all the correlations a positive value, we multiple a "-1"to Cd and Zn. In this part we run two different scenarios:
\begin{itemize}
  \item[Scenario I.] Selecting Pb and Zn as the two main variables with high correlation ($0.60$) and Cd and S as the two main variables with low correlation ($0.06$). In this scenario we used perfect ranking, and we didn't use auxiliary variables.
  \item[Scenario II.] From a chemical point of view we, selecting Cd and Pb as the two main variables and for sorting them using two auxiliary variables; Zn with $0.48$ correlation with Cd and S with $0.27$ with Pb. This is a heuristic approach. Basically economical or sociological information or the density of highways could also serve as auxiliary variables.
\end{itemize}

Results are shown in table \ref{tab15} (Scenario I) and table \ref{tab16} (Scenario II). In table \ref{tab15}, efficiency of estimators for estimating the means of Pb and Zn with 0.6 correlation, and the means of Cd and S with 0.06 correlation are presented. For two variables with reasonable correlation (Zn and Pb) MVSR is not bad, because ranking just based on the first variable, supports the second variable.\\
For Cd and S, the situation is worse for MVSR, because of weak correlation around 0.06 between them. The first variable is not able to support the second one. Efficiency for S in MVSR is around 1. But for CPOR and RPOR results for the second variable are better. With decreasing efficiency of the first variable (Cd) from MVRS to CPOR and RPOR, the efficiency of the second variable (S) raise reasonably. Average of efficiency for S in MVSR is around 1.01 but in CPOR and RPOR are around 1.09. Again, $K/n$ is the most important parameter in efficiency and after that $m$.\\\\

In table \ref{tab16}, we have used two auxiliary variables to rank the main variables. For Zn we have used Cd with 0.48 and for Pb we have used S with 0.27 correlations. As we can see, MVRS just improves efficiency of the first variable (Cd) and CPOR and RPOR improve the both variables estimations however the improvement is not so large because of almost week correlations between auxiliary variables and the main variables ($0.48$ and $0.27$). \\\\\\
Also table \ref{tab17} presents Monte Carlo expectation of the estimators that shows unbiasness of the estimators.\\
 By our sampling technique mean values referring to a complete set of 59 geographical units are obtained. Clearly the regional relation is not taken into regard (which is already done by papers mentioned above) but there is now a number available which can characterize the status of Baden-Württemberg overall, and for example a time series could be done to see the general changes with respect to the pollution.

\section{Conclusion}
CPOR and RPOR can be used for implement RSS in population surveys where
there are multiple variables of interest. CPOR and RPOR enhance the
parameters estimation simultaneously with a reasonable sample size, that most of the RSS methods can not do in multiple
variables cases. As we see in the real case studies, for CPOR and ROPR
there are no need to use perfect ranking using the main variables
and it can be done using some variables, easy to measure, with
 reasonable correlation with the main variables.
 The simulation section and real case study confirmed the assertions in the paper.\\
For further works, it would be beneficial to find some unbiased estimators for
variance of CPOR. Because of randomness of $K_h$ it is not easy to calculate variance and an unbiased estimator of variance for CPOR but as CPOR uses information of all LEs and in simulations we saw that CPOR was more efficient than RPOR in almost all the cases and maybe it is reasonable to use variance estimator of RPOR as a conservative estimate for variance of CPOR.\\\\\\
% and then it would be interesting to find a design for
%ranking using Posets in which $K_h$'s are all equal and they are
%not random variable any more. For example we can instead to use mean rank of the elements in
%LEs, select one of the LEs, for each set (for example select one
%of the column in table 4), and according to order of that LE, put
%the elements in the strata. Then here size of all strata are $K$
%and they are not random variable any more.

\large{\bf Appendix A.
Proving Theorem 1}\\\\

Proof of the theorem is the same as Panahbehagh et al. (2017) and just please note that here $E(I_{[h]i})=E(I_{(h)i})$.\\
\\\\\\\\

\large{\bf Appendix B.
Proving Theorem 2}\\\\
Here according to the sampling strategy, (i) taking an $iid$ sample from $f$ (a model) and (ii) taking an stratified finite population sampling from the selected sample (a design), we have a Model-Design based sampling, let indexes of $"M"$ and $"D"$, mean "according to the
Model and the Design" respectively.
Then with

\begin{equation*}
\widehat{\mu}_{P}^{j}=\sum\limits_{h=1}^{m}W_{h}\bar{X}_{\{h\}}^{j}
\end{equation*}
where%
\begin{eqnarray*}
I_{\{h\}i}=\left\{\begin{matrix}
1
 & \;\;\;\;  \text{if $X^{j}_{\{h\}i}$ is in the $s_{\{h\}}$}
 &\\\\
0 & \;\;\;\;  \text{otherwise}
 &
\end{matrix}\right.
\end{eqnarray*}
we have
\begin{eqnarray*}
E(\widehat{\mu }_{P}^{j}) &=&E_{M}[E_{D}(\widehat{\mu }_{P}^{j}|%
\mathbf{X}_{Km})]=E_{M}[\sum\limits_{h=1}^{m}\frac{K_{h}}{Km}\frac{1}{n_{h}}%
\sum_{i=1}^{K_{h}}X_{\{h\}i}^{j}E_{D}(I_{\{h\}i}|\mathbf{X}_{Km})] \\
&=&E_{M}(\sum\limits_{h=1}^{m}\frac{K_{h}}{Km}\frac{1}{n_{h}}\sum_{i=1}^{K_{h}}X_{%
\{h\}i}^{j}\frac{n_{h}}{K_{h}})=\frac{1}{Km}E_{M}(\sum\limits_{h=1}^{m}%
\sum\limits_{i=1}^{K}X_{hi}^{j})=\mu^{j}
\end{eqnarray*}
where $\mathbf{X}_{Km}$ indicates whole sample of size Km.\\
\\\\\\\\

\large{\bf Appendix C.
Proving Theorem 3}\\\\
Here the design affected by two sources of variations; variation from selecting one of the LEs and variation from selection the sample from the fixed form of the stratified population conditional on the result of the LEs which we indicate them with $"D_1"$ and $"D_2"$ respectively. Therefore here based on LEs assume we have $\mathbf{X}_{Km.q}; q=1,2,...,Q$ and $\mathbf{X}_{Km.q}$ may happen with probability $L_q$. Please note that $L_q=\frac{1}{Q}$ because all combinations of LEs happen with equal probability. Then we have
\begin{equation*}
E(\widehat{\mu}_{R}^{j})=E_ME_{D_1}E_{D_2}(\widehat{\mu}_{R}^{j})
\end{equation*}
now as
\begin{equation*}
E_{D_2}(I_{[h\}i})= \frac{n}{K}
\end{equation*}
and
\begin{equation*}
E_{D_1}E_{D_2}(\widehat{\mu}_{R}^{j})=E_{D_1}(\bar{X}_{Km})=\bar{X}_{Km}
\end{equation*}
we have
\begin{equation*}
E(\widehat{\mu}_{R}^{j})=E_M[\frac{1}{mK}\sum\limits_{h=1}^{m}\sum\limits_{i=1}^{K}X^j_{hi}]=\mu^j
\end{equation*}
then $E(\widehat{\mu}_{R}^{j})=\mu_{R}^{j}$.\\\\
For variance we have
\begin{equation*}
V(\widehat{\mu}^j_{R})=V_ME_{D_1}E_{D_2}(\widehat{\mu }^j_{R})+E_MV_{D_1}E_{D_2}(\widehat{\mu }^j_{R})+E_ME_{D_1}V_{D_2}(\widehat{\mu }^j_{R}).
\end{equation*}
It is easy to see that
\begin{equation*}
V_{M}E_{D_1}E_{D_2}(\widehat{\mu }^j_{V})=\frac{\sigma^2_j}{Km}
\end{equation*}

and then as $V_{D_1}E_{D_2}(\widehat{\mu }^j_{R})=0$ (because $E_{D_2}(\widehat{\mu }^j_{R})=\bar{X}_{Km}$ is not variable respect to $D_1$) we have
\begin{equation*}
E_{D_1}V_{D_2}(\widehat{\mu}^j_{R})=\frac{1}{m^2}\sum\limits_{h=1}^m\frac{1-\frac{n}{K}}{n}\frac{1}{Q}\sum\limits_{q=1}^QS^2_{[h\}qjK}
\end{equation*}

and therefore
\begin{equation*}
V(\widehat{\mu}^j_{R})=\frac{\sigma^2_j}{Km}+\frac{1}{m^2}\sum\limits_{h=1}^m\frac{1-\frac{n}{K}}{n}E_M(\frac{1}{Q}\sum\limits_{q=1}^QS^2_{[h\}qjK}).
\end{equation*}
\\\\

For the unbiased estimator of the variance first note that
%\begin{equation*}
%E(s^2_{[h\}j})=E_ME_{D_1}E_{D_2}(s^2_{[h\}j})=E_ME_{D_1}(S^2_{[h\}j})=E_M(\frac{1}{Q}\sum\limits_{q=1}^QS^2_{[h\}qjK}),
%\end{equation*}
%then
%\begin{equation*}
%\frac{1}{m^2}\sum\limits_{h=1}^m\frac{1-\frac{n}{K}}{n}s^2_{[h\}j}
%\end{equation*}
%is unbiased for the second part of \ref{VR}. For the first part of \ref{VR} we should estimate $\sigma^2_j$ unbiasedly. Here
as we take an iid sample for each set and rank them in $m$ ranks then rank for each unit is distributed uniformly in vector $(1,2,...,m)$ and therefore we have

\begin{equation*}\label{muord}
\mu^j=E(X^j_1)=EE(X^j_1|rank(X^j_1))=\frac{1}{m}\sum\limits_{h=1}^{m}E(X^j_1|rank(X^j_1)=h)=\frac{1}{m}\sum\limits_{h=1}^{m}\mu^j_{[h\}},
\end{equation*}
\begin{eqnarray}\label{varord}
\sigma^2_j&=&VE(X^j|rank(X^j))+EV(X^j|rank(X^j))\\ \nonumber
&=&V[\sum\limits_{h=1}^{m}\mu^j_{[h\}} I(rank(X^j)=h)]+E[\sum\limits_{h=1}^{m}\sigma_{[h\}j}^2 I(rank(X^j)=h)]\\ \nonumber
&=& \frac{1}{m}\sum\limits_{h=1}^{m}(\mu^j_{[h\}}-\mu)^2+\frac{1}{m}\sum\limits_{h=1}^{m}\sigma^2_{[h\}j}
\end{eqnarray}
where $rank(X_1)$ indicates rank of $X_1$ in its selected set and $I(rank(X^j_1)=h)$ is an indicator function which takes 1, if $rank(X^j_1)=h$.\\
 Then
\begin{eqnarray*}\label{muord}
E(\widehat{V}(\widehat{\mu}^j_{R}))=\frac{1}{nm(Km-1)}[E(\sum\limits_{h=1}^{m}\sum_{i\epsilon s_{[h\}}}(X^j_{[h\}i}-\widehat{\mu}_{R}^{j})^2)+(K-n)E(\sum\limits_{h=1}^{m}s^2_{[h\}j})].
\end{eqnarray*}
Now as
\begin{eqnarray*}
E(\sum\limits_{h=1}^{m}s^2_{[h\}j})=\sum\limits_{h=1}^mE_M(\frac{1}{Q}\sum\limits_{q=1}^Q S^2_{[h\}}qjK),
\end{eqnarray*}
and
\begin{eqnarray*}
&&E(\sum\limits_{h=1}^{m}\sum_{i\epsilon s_{[h\}}}(X^j_{[h\}i}-\widehat{\mu}_{R}^{j})^2)\\
&=&E(\sum\limits_{h=1}^{m}\sum_{i\epsilon s_{[h\}}}(X^j_{[h\}i})^2)-nmE(\widehat{\mu}_{R}^{j})^2\\
&=&E(\sum\limits_{h=1}^{m}\sum_{i=1}^K(X^j_{[h\}i})^2I_{[h\}i})-nmV(\widehat{\mu}_{R}^{j})-nmE^2(\widehat{\mu}_{R}^{j})\\
&=&\sum\limits_{h=1}^{m}\frac{n}{K}K(\sigma^2_{[h\}j}+(\mu^j_{[h\}})^2)-\frac{n}{K}\sigma^2_j-\frac{1}{m}\sum\limits_{h=1}^m(1-\frac{n}{K})E_M(\frac{1}{Q}\sum\limits_{q=1}^QS^2_{[h\}qjK})-nm\mu^2\\
&=&nm(\frac{1}{m}(\sum\limits_{h=1}^{m}\sigma^2_{[h\}j}+\sum\limits_{h=1}^{m}(\mu^j_{[h\}}-\mu)^2))-\frac{n}{K}\sigma^2_j-\frac{1}{m}\sum\limits_{h=1}^m(1-\frac{n}{K})E_M(\frac{1}{Q}\sum\limits_{q=1}^QS^2_{[h\}qjK})\\
&=&(nm-\frac{n}{K})\sigma^2_j-\frac{1}{m}\sum\limits_{h=1}^m (1-\frac{n}{K})E_M(\frac{1}{Q}\sum\limits_{q=1}^Q S^2_{[h\}qjK})
\end{eqnarray*}
where the last equation is based on \ref{varord}, we have

\begin{equation*}
E(\widehat{V}(\widehat{\mu}^j_{R}))=V(\widehat{\mu}^j_{R})
\end{equation*}

%\large{\bf Appendix C.}\\\\

%\begin{eqnarray*}
%E(\sum\limits_{h=1}^{m}\frac{1}{n_h-1}\sum_{i\epsilon s_{\{h\}}}(X^j_{\{h\}i}-
%\bar{X}^j_{\{h\}})^2)=E(\sum\limits_{h=1}^{m}E(\frac{1}{n_h-1}[\sum_{i=1}^{K_h}
%(X^j_{\{h\}i})^2I_{\{h\}i}-n_h(\bar{X}^j_{\{h\}})^2]|K_h))\\
 %= %E(\sum\limits_{h=1}^{m}\frac{1}{n_h-1}[\sum_{i=1}^{K_h}\frac{n_h}{K_h}(Var({X}^j_{\{h\}i})+E^2({X}^j_{\{h\}i}))-n_h(Var(\bar{X}^j_{\{h\}})+E^2(\bar{X}^j_{\{h\}}))] % ) \\
%=E(\sum\limits_{h=1}^{m}\frac{1}{n_h-1}[n_h(\sigma_{\{h\}j}^2+(\mu^j_{\{h\}})^2)-n_h(\frac{1-n_h/K_h}{n_h}\sigma_{\{h\}j}^2+\frac{1}{K_h}\sigma_{\{h\}j}^2+(\mu^j_{\{h\}})^2)]\\
%=E(\sum\limits_{h=1}^{m}\frac{1}{n_h-1}(n_h-1)\sigma_{\{h\}j}^2)=\sum\limits_{h=1}^{m}\sigma_{\{h\}j}^2
%\end{eqnarray*}
\newpage
\begin{small}
{}
\end{small}

\newpage
\begin{table}[h]
\begin{center}
\caption{ Virtual strata, using conventional RSS.\label{tab2}}
\end{center}
\begin{center}
\begin{tabular}{c|c|c|c}
 $1^{st}$ stratum & $2^{nd}$ stratum & $\cdots $ & $m^{th}$
stratum
\\ \hline
$\textbf{X}_{(1)1}$ & $\textbf{X}_{(2)1}$ & $\cdots $ & $\textbf{X}_{(m)1}$ \\
$\textbf{X}_{(1)2}$ & $\textbf{X}_{(2)2}$ & $\cdots $ & $\textbf{X}_{(m)2}$ \\
$\vdots $ & $\vdots $ & $\ddots $ & $\vdots $ \\
$\textbf{X}_{(1)K}$ & $\textbf{X}_{(2)K}$  &$\cdots $& $\textbf{X}_{(m)K}$ \\
\end{tabular}%
\end{center}
\end{table}

\begin{table}[h]
\begin{center}
\caption{ Elements of a set with their variables.\label{tab4}}
\end{center}
\begin{center}
\begin{tabular}{c|cc}
& $X^{1}$ & $X^{2}$ \\ \hline
a & 0 & 1 \\
b & 2 & 1 \\
c & 1 & 2 \\
d & 3 & 3 \\
e & 0 & 4%
\end{tabular}%
\end{center}
\end{table}

\begin{table}[h]
\begin{center}
\caption{All possible LEs with respect to Posets.\label{tab5}}
\end{center}
\begin{center}
\begin{tabular}{c|c|c|c|c|c|c|c}
LE1 & LE2 & LE3 & LE4 & LE5 & LE6 & LE7 & LE8 \\ \hline
d & d & d & e & d & d & d & e \\
c & c & e & d & b & b & e & d \\
b & e & c & c & c & e & b & b \\
e & b & b & b & e & c & c & c \\
a & a & a & a & a & a & a & a \\
\end{tabular}%
\end{center}
\end{table}

\begin{table}[h]
\begin{center}
\caption{ Mean height of each element in all possible LEs.\label{tab6}}
\end{center}
\begin{center}
\begin{tabular}{c|c|c}
& mean height & rounded height\\ \hline
a & 1 &1\\
b & 2.875&3 \\
c & 2.875 &3\\
d & 4.75 &5\\
e & 3.5 &4\\
\end{tabular}%
\end{center}
\end{table}

\begin{table}[h]
\begin{center}
\caption{ Putting the elements of a set in strata.\label{tab7}}
\end{center}
\begin{center}
\begin{tabular}{c|c|c|c|c|c}
 strata & 1 & 2 & 3 & 4 & 5 \\ \hline
& a & & b & e & d \\
& & & c & & \\
\end{tabular}%
\end{center}
\end{table}

\begin{table}[h]
\begin{center}
\caption{ Virtual strata, using Posets ranking.\label{tab8}}
\end{center}
\begin{center}
\begin{tabular}{c|c|c|c}
 $1^{st}$ stratum & $2^{nd}$ stratum & $\cdots $ & $m^{th}$
stratum
\\ \hline
$\textbf{X}_{\{1\}1}$ & $\textbf{X}_{\{2\}1}$ & $\cdots $ & $\textbf{X}_{\{m\}1}$ \\
$\textbf{X}_{\{1\}2}$ & $\textbf{X}_{\{2\}2}$ & $\cdots $ & $\textbf{X}_{\{m\}2}$ \\
$\vdots $ & $\vdots $ & $\ddots $ & $\vdots $ \\
$\vdots $ & $\vdots $ & $\cdots $ & $\textbf{X}_{\{m\}K_{m}}$ \\
$\textbf{X}_{\{1\}K_{1}}$ & $\vdots $ & & \\
& $\textbf{X}_{\{2\}K_{2}}$ & & \\
\end{tabular}%
\end{center}
\end{table}

\begin{table}[h]
\begin{center}
\caption{Efficiency of the estimators in bivariate normal case $(X^{1},X^{2})\sim B.N(0,0,1,1,\rho)$ with solving problem of negative correlation.\label{tab12}}
\end{center}
\begin{center}
\begin{tabular}{c|c|c|c|c|c}
&$\rho$=-0.9&$\rho$=-0.5&$\rho$=0&$\rho$=-0.5$\rightarrow$0.5&$\rho$=-0.9$\rightarrow$0.9\\
\hline
$\widehat{\mu}_{V}^{1}$&1.32&1.32 &1.33&1.30&1.31\\
$\widehat{\mu}_{P}^{1}$&1.02&1.02&1.07&1.21& 1.31\\
$\widehat{\mu}_{R}^{1}$&0.99&1.02&1.04&1.16& 1.28\\\hline
$\widehat{\mu}_{V}^{2}$&1.26 &1.08 &1.01&1.10&1.25\\
$\widehat{\mu}_{P}^{2}$&1.02&1.00&1.06&1.22&1.31\\
$\widehat{\mu}_{R}^{2}$&0.99&1.02&1.04&1.16&1.28\\
\end{tabular}%
\end{center}
\end{table}

\begin{table}[h]
\begin{center}
\caption{Efficiency of the estimators for different cases for bivariate normal distribution.\label{tab13}}
\end{center}
\begin{center}
\begin{tabular}{cccc||c|ccc}

$m$ & $K$ & $n$ & $\rho$ & variable & $\widehat{\mu}_{V}$
& $\widehat{\mu}_{P}$ &$\widehat{\mu}_{R}$ \\ \hline\hline
\hline
3 & 12 & 4 & 0.3 & $X^1$ & 1.49 & 1.16 & 1.12  \\
& & & & $X^2$ & 1.00 & 1.12 & 1.11  \\ \hline
& & & 0.5 & $X^1$& 1.45 & 1.20 &1.17  \\
& & & & $X^2$ & 1.05 & 1.21&1.18  \\ \hline
& & & 0.7 & $X^1$& 1.47 & 1.27 &1.26 \\
& & & & $X^2$ & 1.17 & 1.30 &1.26 \\ \hline
& & & 0.9 & $X^1$& 1.49 & 1.41 & 1.39\\
& & & & $X^2$ & 1.35 & 1.42 &1.41 \\ \hline\hline \hline
& & 6 & 0.3 & $X^1$ & 1.31 & 1.13&1.10  \\
& & & & $X^2$ & 1.01 & 1.10 &1.07\\ \hline
& & & 0.5 & $X^1$& 1.30 & 1.16 & 1.13 \\
& & & & $X^2$ & 1.05 & 1.14 &1.11 \\ \hline
& & & 0.7 &$X^1$ & 1.33 & 1.23 &1.19  \\
& & & & $X^2$ & 1.13 & 1.23 &1.20  \\ \hline
& & & 0.9 &$X^1$ & 1.32 & 1.31 &1.29 \\
& & & & $X^2$ & 1.23 & 1.31 &1.27 \\
\end{tabular}%
\end{center}
\end{table}

\begin{table}[h]
\begin{center}
\caption{Efficiency of the estimators for estimating the means of Fdw and Esn as the main variables and Fht and Npl as the auxiliary variables with $0.78$ and $0.71$ correlations. \label{tab14}}
\begin{tabular}{ccc||cc|cc|cc}
	&		&		&	$\widehat{\mu}_{V}^{1}$	&	$\widehat{\mu}_{V}^{2}$	&	$\widehat{\mu}_{P}^{1}$	&	$\widehat{\mu}_{P}^{2}$	&	$\widehat{\mu}_{R}^{1}$	&	$\widehat{\mu}_{R}^{2}$	\\	
K	&	m	&	n	&	Fdw	&	Esn	&	Fdw	&	Esn	&	Fdw	&	Esn	\\	\hline
5	&	3	&	2	&	1.40	&	1.11	&	1.32	&	1.17	&	1.28	&	1.14	\\	
	&		&	3	&	1.23	&	1.07	&	1.18	&	1.06	&	1.18	&	1.07	\\	
	&		&	4	&	1.10	&	1.04	&	1.09	&	1.04	&	1.09	&	1.05	\\	
	&	5	&	2	&	1.63	&	1.18	&	1.45	&	1.24	&	1.45	&	1.23	\\	
	&		&	3	&	1.35	&	1.10	&	1.26	&	1.11	&	1.24	&	1.13	\\	
	&		&	4	&	1.14	&	1.05	&	1.11	&	1.06	&	1.11	&	1.06	\\	
	&	7	&	2	&	1.77	&	1.19	&	1.55	&	1.27	&	1.53	&	1.27	\\	
	&		&	3	&	1.40	&	1.10	&	1.29	&	1.12	&	1.30	&	1.14	\\	
	&		&	4	&	1.17	&	1.06	&	1.13	&	1.07	&	1.13	&	1.07	\\	\hline
7	&	3	&	3	&	1.36	&	1.09	&	1.26	&	1.15	&	1.25	&	1.13	\\	
	&		&	5	&	1.15	&	1.06	&	1.13	&	1.07	&	1.12	&	1.08	\\	
	&		&	6	&	1.09	&	1.03	&	1.07	&	1.03	&	1.06	&	1.03	\\	
	&	5	&	3	&	1.58	&	1.16	&	1.43	&	1.22	&	1.40	&	1.20	\\	
	&		&	5	&	1.23	&	1.07	&	1.18	&	1.08	&	1.17	&	1.09	\\	
	&		&	6	&	1.10	&	1.03	&	1.08	&	1.04	&	1.08	&	1.04	\\	
	&	7	&	3	&	1.71	 &	1.19	&	1.53	&	1.25	&	1.51	&	1.24	\\	
	&		&	5	&	1.26	&	1.08	&	1.22	&	1.09	&	1.20	&	1.11	\\	
	&		&	6	&	1.12	&	1.04	&	1.09	&	1.05	&	1.09	&	1.05	\\	\hline \hline
	&	Average	&		&	1.32	&	1.09	&	1.24	&	1.12	&	1.23	&	1.12	\\	

\end{tabular}%
\end{center}
\end{table}
\begin{table}[h]
\begin{center}
\caption{ Efficiency of the estimators for estimating the means of Pb and Zn with $0.6$ correlation and the means of Cd and S with $0.06$ correlation. Here we used complete ranking. \label{tab15}}
\begin{tabular}{ccc|||cc|cc|cc||cc|cc|cc}
	&		&		&	$\widehat{\mu}_{V}^{1}$	&	$\widehat{\mu}_{V}^{2}$	&	$\widehat{\mu}_{P}^{1}$	&	$\widehat{\mu}_{P}^{2}$	&	$\widehat{\mu}_{R}^{1}$	&	$\widehat{\mu}_{R}^{2}$	&	$\widehat{\mu}_{V}^{1}$	&	$\widehat{\mu}_{V}^{2}$	&	$\widehat{\mu}_{P}^{1}$	&	$\widehat{\mu}_{P}^{2}$	&	$\widehat{\mu}_{R}^{1}$	&	$\widehat{\mu}_{R}^{2}$	\\	
m	&	K	&	n	&	Pb	&	Zn	&	Pb	&	Zn	&	Pb	&	Zn	&	Cd	&	S	&	Cd	&	S	&	Cd	&	S	\\	\hline
3	&	5	&	2	&	1.32	&	1.11	&	1.13	&	1.21	&	1.12	&	1.15	&	1.36	&	1.01	&	1.13	&	1.09	&	1.12	&	1.09	\\	
	&		&	4	&	1.11	&	1.02	&	1.06	&	1.03	&	1.05	&	1.03	&	1.11	&	1.01	&	1.05	&	1.00	&	1.05	&	1.03	\\	
	&	7	&	2	&	1.41	&	1.11	&	1.15	&	1.16	&	1.10	&	1.12	&	1.41	&	0.99	&	1.16	&	1.08	&	1.11	&	1.07	\\	
	&		&	4	&	1.25	&	1.08	&	1.11	&	1.10	&	1.09	&	1.10	&	1.16	&	1.01	&	1.00	&	1.01	&	1.03	&	1.06	\\	
	&	10	&	2	&	1.59	&	1.16	&	1.24	&	1.25	&	1.16	&	1.17	&	1.53	&	1.04	&	1.22	&	1.12	&	1.19	&	1.11	\\	
	&		&	4	&	1.38	&	1.11	&	1.15	&	1.17	&	1.12	&	1.14	&	1.29	&	1.02	&	1.16	&	1.09	&	1.07	&	1.07	\\	\hline
5	&	5	&	2	&	1.61	&	1.21	&	1.23	&	1.27	&	1.18	&	1.23	&	1.53	&	1.02	&	1.19	&	1.12	&	1.15	&	1.10	\\	
	&		&	4	&	1.15	&	1.04	&	1.05	&	1.05	&	1.04	&	1.06	&	1.13	&	1.01	&	1.05	&	1.02	&	1.04	&	1.02	\\	
	&	7	&	2	&	1.69	&	1.21	&	1.23	&	1.30	&	1.21	&	1.26	&	1.62	&	1.00	&	1.23	&	1.11	&	1.18	&	1.08	\\	
	&		&	4	&	1.35	&	1.14	&	1.14	&	1.15	&	1.13	&	1.16	&	1.33	&	1.00	&	1.12	&	1.04	&	1.09	&	1.04	\\	
	&	10	&	2	&	1.93	&	1.28	&	1.31	&	1.37	&	1.24	&	1.34	&	1.78	&	0.99	&	1.28	&	1.13	&	1.21	&	1.12	\\	
	&		&	4	&	1.56	&	1.21	&	1.20	&	1.28	&	1.17	&	1.26	&	1.47	&	1.03	&	1.15	&	1.12	&	1.13	&	1.11	\\	\hline
7	&	5	&	2	&	1.69	&	1.21	&	1.26	&	1.31	&	1.21	&	1.28	&	1.63	&	1.04	&	1.26	&	1.18	&	1.21	&	1.19	\\	
	&		&	4	&	1.16	&	1.10	&	1.09	&	1.10	&	1.07	&	1.12	&	1.15	&	0.99	&	1.06	&	1.04	&	1.06	&	1.01	\\	
	&	7	&	2	&	1.90	&	1.28	&	1.27	&	1.35	&	1.29	&	1.32	&	1.85	&	1.03	&	1.30	&	1.12	&	1.28	&	1.09	\\	
	&		&	4	&	1.45	 &	1.19	&	1.19	&	1.19	&	1.17	&	1.20	&	1.37	 &	1.02	&	1.17	&	1.07	&	1.17	&	1.07	\\	
	&	10	&	2	&	2.20	&	1.36	&	1.40	&	1.49	&	1.39	&	1.49	&	1.99	&	1.02	&	1.40	&	1.19	&	1.31	&	1.15	\\	
	&		&	4	&	1.66	&	1.30	&	1.27	&	1.36	&	1.25	&	1.34	&	1.62	&	1.02	&	1.25	&	1.13	&	1.24	&	1.12	\\	\hline \hline
	&	Average	&		&	1.52	&	1.17	&	1.19	&	1.23	&	1.17	&	1.21	&	1.46	&	1.01	&	1.18	&	1.09	&	1.15	&	1.09	\\	
\end{tabular}%
\end{center}
\end{table}
\begin{table}[h]
\begin{center}
\caption{Efficiency of the estimators for estimating the means of Cd and Pb as the main variables and Zn and S as the auxiliary variables with $0.48$ and $0.27$ correlations. \label{tab16}}
\begin{tabular}{ccc||cc|cc|cc}
	&		&		&	$\widehat{\mu}_{V}^{1}$	&	$\widehat{\mu}_{V}^{2}$	&	$\widehat{\mu}_{P}^{1}$	&	$\widehat{\mu}_{P}^{2}$	&	$\widehat{\mu}_{R}^{1}$	&	$\widehat{\mu}_{R}^{2}$	\\	
m	&	K	&	n	&	Cd	&	Pb	&	Cd	&	Pb	&	Cd	&	Pb	\\	\hline
3	&	5	&	2	&	1.31	&	1.01	&	1.07	&	1.03	&	1.02	&	1.02	\\	
	&		&	4	&	1.08	&	1.01	&	0.99	&	0.98	&	1.00	&	1.01	\\	
	&	7	&	2	&	1.40	&	0.99	&	1.02	&	1.01	&	1.04	&	1.03	\\	
	&		&	4	&	1.20	&	1.00	&	1.00	&	1.00	&	1.02	&	1.01	\\	
	&	10	&	2	&	1.46	&	0.99	&	1.04	&	1.02	&	1.03	&	1.03	\\	
	&		&	4	&	1.31	&	0.99	&	1.02	&	1.03	&	1.01	&	1.01	\\	\hline
5	&	5	&	2	&	1.48	&	1.00	&	1.07	&	1.05	&	1.04	&	1.05	\\	
	&		&	4	&	1.13	&	1.00	&	1.00	&	1.00	&	1.01	&	1.01	\\	
	&	7	&	2	&	1.63	&	1.00	&	1.03	&	1.05	&	1.05	&	1.04	\\	
	&		&	4	&	1.31	&	1.00	&	1.01	&	1.04	&	1.03	&	1.04	\\	
	&	10	&	2	&	1.82	&	1.01	&	1.06	&	1.09	&	1.07	&	1.07	\\	
	&		&	4	&	1.49	&	1.00	&	1.07	&	1.06	&	1.04	&	1.04	\\	\hline
7	&	5	&	2	&	1.62	&	1.02	&	1.07	&	1.10	&	1.06	&	1.08	\\	
	&		&	4	&	1.14	&	1.00	&	1.01	&	1.02	&	1.02	&	1.02	\\	
	&	7	&	2	&	1.85	&	1.01	&	1.07	&	1.07	&	1.08	&	1.09	\\	
	&		&	4	&	1.37	 &	1.01	&	1.03	&	1.04	&	1.05	&	1.05	\\	
	&	10	&	2	&	2.05	&	1.01	&	1.09	&	1.08	&	1.08	&	1.07	\\	
	&		&	4	&	1.61	&	1.02	&	1.07	&	1.10	&	1.06	&	1.07	\\	\hline \hline
	&	Average	&		&	1.46	&	1.00	&	1.04	&	1.04	&	1.04	&	1.04	\\	

\end{tabular}%
\end{center}
\end{table}

\begin{table}[h]
\begin{center}
\caption{Expectation of the estimators based on Scenario I. \label{tab17}}
\begin{tabular}{ccc|cccccc|cccccc}
	&		&		&	$\widehat{\mu}_{V}^{1}$	&	$\widehat{\mu}_{V}^{2}$	&	$\widehat{\mu}_{P}^{1}$	&	$\widehat{\mu}_{P}^{2}$	&	$\widehat{\mu}_{R}^{1}$	&	$\widehat{\mu}_{R}^{2}$	&	$\widehat{\mu}_{V}^{1}$	&	$\widehat{\mu}_{V}^{2}$	&	$\widehat{\mu}_{P}^{1}$	&	$\widehat{\mu}_{P}^{2}$	&	$\widehat{\mu}_{R}^{1}$	&	$\widehat{\mu}_{R}^{2}$	\\	
m	&	K	&	n	&	Pb	& Zn	&	Pb	&	Zn	&	Pb	&	Zn	&	Cd	&	S	&	Cd	&	S	&	Cd	&	S	\\	\hline
3	&	5	&	2	&	0.9	&	132.9	&	0.9	&	132.3	&	0.9	&	132.3	&	0.1	&	1787.9	&	0.1	&	1786.0	&	0.1	&	1789.7	\\	
	&		&	4	&	0.9	&	133.7	&	0.9	&	133.4	&	0.9	&	131.5	&	0.1	&	1789.7	&	0.1	&	1789.5	&	0.1	&	1788.3	\\	
	&	7	&	2	&	0.9	&	134.1	&	0.9	&	132.2	&	0.9	&	134.1	&	0.1	&	1787.7	&	0.1	&	1788.4	&	0.1	&	1787.0	\\	
	&		&	4	&	0.9	&	132.2	&	0.9	&	132.3	&	0.9	&	132.7	&	0.1	&	1787.9	&	0.1	&	1787.3	&	0.1	&	1788.8	\\	
	&	10	&	2	&	0.9	&	132.0	&	0.9	&	132.8	&	0.9	&	133.7	&	0.1	&	1783.6	&	0.1	&	1789.3	&	0.1	&	1786.2	\\	
	&		&	4	&	0.9	&	132.4	&	0.9	&	132.8	&	0.9	&	132.4	&	0.1	&	1788.4	&	0.1	&	1786.3	&	0.1	&	1785.3	\\	\hline
5	&	5	&	2	&	0.9	&	133.3	&	0.9	&	133.3	&	0.9	&	133.8	&	0.1	&	1788.2	&	0.1	&	1788.7	&	0.1	&	1787.2	\\	
	&		&	4	&	0.9	&	132.7	&	0.9	&	133.0	&	0.9	&	133.1	&	0.1	&	1787.7	&	0.1	&	1787.7	&	0.1	&	1787.1	\\	
	&	7	&	2	&	0.9	&	133.3	&	0.9	&	132.8	&	0.9	&	133.1	&	0.1	&	1788.8	&	0.1	&	1787.4	&	0.1	&	1786.1	\\	
	&		&	4	&	0.9	&	133.5	&	0.9	&	133.5	&	0.9	&	132.5	&	0.1	&	1788.7	&	0.1	&	1788.5	&	0.1	&	1788.1	\\	
	&	10	&	2	&	0.9	&	133.8	&	0.9	&	132.9	&	0.9	&	131.9	&	0.1	&	1790.0	&	0.1	&	1787.9	&	0.1	&	1782.9	\\	
	&		&	4	&	0.9	&	132.7	&	0.9	&	132.8	&	0.9	&	132.8	&	0.1	&	1789.0	&	0.1	&	1787.8	&	0.1	&	1787.8	\\	\hline
7	&	5	&	2	&	0.9	&	133.0	&	0.9	&	133.1	&	0.9	&	133.4	&	0.1	&	1788.4	&	0.1	&	1789.7	&	0.1	&	1787.7	\\	
	&		&	4	&	0.9	&	132.6	&	0.9	&	132.7	&	0.9	&	132.4	&	0.1	&	1788.9	&	0.1	&	1789.2	&	0.1	&	1787.2	\\	
	&	7	&	2	&	0.9	&	132.9	&	0.9	&	133.3	&	0.9	&	132.8	&	0.1	&	1787.2	&	0.1	&	1786.6	&	0.1	&	1785.7	\\	
	&		&	4	&	0.9	 &	133.2	&	0.9	&	133.0	&	0.9	&	133.1	&	0.1	 &	1786.0	&	0.1	&	1787.4	&	0.1	&	1787.0	\\	
	&	10	&	2	&	0.9	&	132.5	&	0.9	&	132.9	&	0.9	&	132.9	&	0.1	&	1787.1	&	0.1	&	1787.3	&	0.1	&	1786.7	\\	
	&		&	4	&	0.9	&	132.9	&	0.9	&	132.8	&	0.9	&	133.2	&	0.1	&	1787.2	&	0.1	&	1786.5	&	0.1	&	1787.8	\\	\hline \hline
	&	Average	&		&	0.9	&	133.0	&	0.9	&	132.9	&	0.9	&	132.9	&	0.1	&	1787.9	&	0.1	&	1787.9	&	0.1	&	1787.0	\\	\hline
	&	Real	&		&	0.9	&	132.9	&	0.9	&	132.9	&	0.9	&	132.9	&	0.1	&	1787.8	&	0.1	&	1787.8	&	0.1	&	1787.8	\\	

\end{tabular}%
\end{center}
\end{table}

%\end{thebibliography}
\end{document}